\documentclass[aip,reprint,jcp,twocolumn]{revtex4-1}
\usepackage{mhchem}
\usepackage[paperheight=11in,paperwidth=8.25in,margin=0.67in]{geometry}
\usepackage{amsmath}
\usepackage{graphicx,color}
\usepackage{url}

\begin{document}
\title{The electronic complexity of the ground-state of the FeMo cofactor of nitrogenase as
relevant to quantum simulations}
\author{Zhendong Li}
\affiliation{Division of Chemistry and Chemical Engineering, California Institute of Technology, Pasadena, CA 91125}
\author{Junhao Li}
\affiliation{Department of Physics, Laboratory of Atomic and Solid-State Physics, Cornell University, Ithaca, NY 14853}
\author{Nikesh S. Dattani}
\affiliation{Oxford University, Hertford College, Oxford, OX1 3BW, UK}
\author{C. J. Umrigar}
\affiliation{Department of Physics, Laboratory of Atomic and Solid-State Physics, Cornell University, Ithaca, NY 14853}
\author{Garnet Kin-Lic Chan}
\affiliation{Division of Chemistry and Chemical Engineering, California Institute of Technology, Pasadena, CA 91125}

\begin{abstract}
  We report that a recent active space model of the nitrogenase FeMo cofactor, proposed
  in the context of quantum simulations, is not
  representative of the electronic structure of the FeMo cofactor ground-state. Although
  quantum resource estimates, outside of the cost of adiabatic state preparation, will not be much affected,
  conclusions should not be drawn from
  the complexity of classical simulations of the electronic structure of this system in this active space. We provide a
  different model active space for the FeMo cofactor that contains the basic open-shell qualitative character, which
  may be useful as a benchmark system for making classical and quantum resource estimates.
\end{abstract}
\maketitle

The process of nitrogen fixation, namely that of converting atmospheric dinitrogen to
a reduced form, such as ammonia, which can then be metabolized by biological species, is essential
to life on this planet\cite{beinert1997iron,howard1996structural,rees2003interface,hoffman2014mechanism}. The industrial Haber-Bosch process to produce fertilizer
from the endothermic reaction ${\rm N_2 + 3H_2 \to 2NH_3}$ is very energy intensive,
requiring a careful balance of high temperatures and high pressures to achieve efficient catalysis.
In contrast, natural bacteria and archaea carry out nitrogen fixation under ambient conditions through nitrogenases.
At the molecular level, the nitrogenase enzyme, an agglomeration of a homodimer Fe protein and the MoFe protein (in its most
common Mo containing form), catalyzes the nitrogen bond-breaking process via a family of 3 metallic cofactors: the
[\ce{Fe4S4}] iron cubane,
the [\ce{Fe8S7}] P cluster, and the [\ce{MoFe7S9C}] FeMo cofactor (FeMoco), with FeMoco serving as the site of nitrogen reduction.
The contrast between the conditions of biological nitrogen fixation and the Haber-Bosch process is an enduring
source of fascination for chemists.

In the search to unravel the secrets of biological nitrogen fixation, the first stage is to understand
the structure of the enzyme itself. After many decades, we now possess atomic scale resolution structures of  nitrogenase, including
all cofactors\cite{spatzal2011evidence,lancaster2011x}. However, the electronic structure of the
cofactors, and in particular the large P cluster and FeMo cofactor, remains poorly understood. This
is due to the complexity of tackling the multiple transition metal ions with their multiple charge states and complicated spin-couplings.
Even though the qualitative electronic structure is believed
to be captured using only the valence active space of the metals and bridging S ligands which provides a great reduction
of the problem size (to, for example, 103 electrons in 71 orbitals in the case of FeMoco considering the Fe 3d, S 3p, Mo 4d, and
the interstitial C 2s2p for the [\ce{MoFe7S9C}] core assuming a total charge $-1$),
no satisfactory classical many-electron
simulation within this valence active space has yet been performed. Because of the need for tangible objectives for quantum simulations
of electronic structure,  these metallic cofactors have thus been suggested as an interesting
target for future quantum simulators. Ref.~\onlinecite{ReiWieSvoWecTro-PNAS-17} provides a pedagogical discussion of the chemical questions that
must be considered when elucidating a complex reaction such as nitrogen fixation, as well as concrete resource estimates
resulting from 54 electron in 54 orbital (54e,54o), and 65 electron in 57 orbital (65e,57o) models of the FeMoco cluster.

\begin{figure}[htb]
  \includegraphics[width=0.4\textwidth]{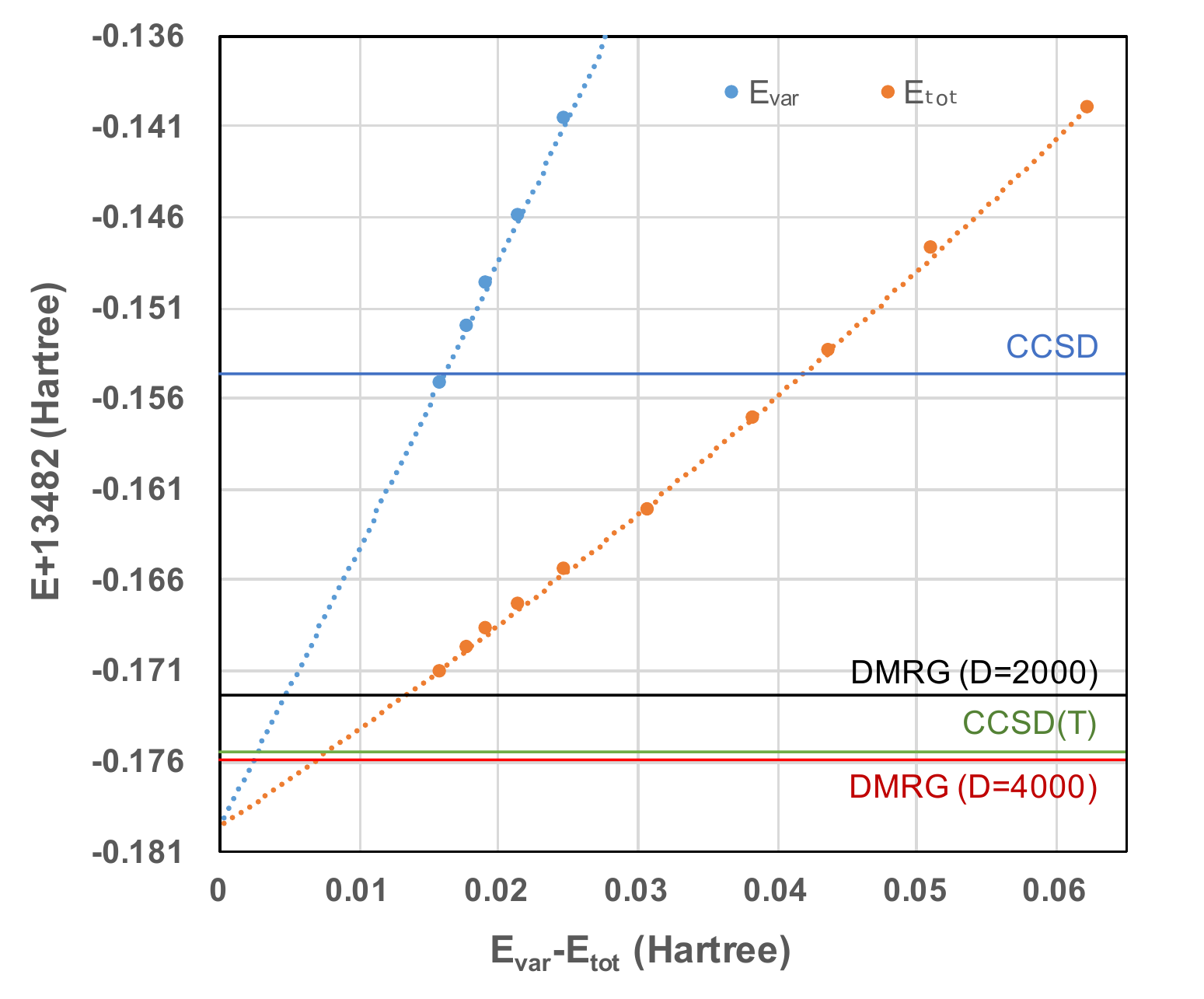}
  \caption{SHCI variational and total energies for progressively decreasing cutoffs (dots) along with quadratic extrapolations (dotted curves)
  of the (54e,54o) model of FeMoco in Ref.~\onlinecite{ReiWieSvoWecTro-PNAS-17}
  (the estimated error in the extrapolated energy is about 2 mHa);
  variational DMRG results at bond dimension $D=2000$ and $D=4000$; CCSD and CCSD(T) energies.
  All calculations are for the $S=0$ state.}
  \label{fig:energies}
\end{figure}

Although the focus of Ref.~\onlinecite{ReiWieSvoWecTro-PNAS-17} was the quantum resource estimates for this problem, it is natural to ask
whether a classical calculation of the electronic structure of FeMoco at the level described in Ref.~\onlinecite{ReiWieSvoWecTro-PNAS-17} is feasible. For this reason,
we report that the active space in Ref.~\onlinecite{ReiWieSvoWecTro-PNAS-17} does not actually contain the representative features of the electronic
structure of the FeMoco ground-state that make its classical simulation difficult. Consequently,
if taken out of context, it provides a misleading characterization
of the classical complexity of obtaining the low-energy states.
In fact, as shown in Fig.~\ref{fig:energies} for the (54e,54o) model of Ref.~\onlinecite{ReiWieSvoWecTro-PNAS-17},
we can obtain accurate ground-state energies ($S=0$)
using standard classical algorithms such as coupled cluster theory~\cite{shavitt2009many},
variational density matrix renormalization group~\cite{white1999ab,ChaHea-JCP-02,chan2011density,ShaCha-JCP-12,schollwock2005density},
and the semistochastic heatbath configuration interaction (SHCI) method~\cite{HolTubUmr-JCTC-16,ShaHolJeaAlaUmr-JCTC-17,LiOttHolShaUmr-ARX-18} (a recent variant of selected configuration interaction plus perturbation theory methods).
The lowest DMRG variational energy and the extrapolated SHCI energy agree to within 5 m$E_h$
or about 0.6 m$E_h$ per metal center (comparable to the 1 m$E_h$ accuracy in
relative energies usually considered to be chemical accuracy).
Note that only modest resources were required for these calculations and higher accuracy, e.g. more variational determinants in selected CI or larger
bond dimensions in DMRG, is very feasible.

\begin{figure}[ht]
\centering
\includegraphics[width=0.4\textwidth]{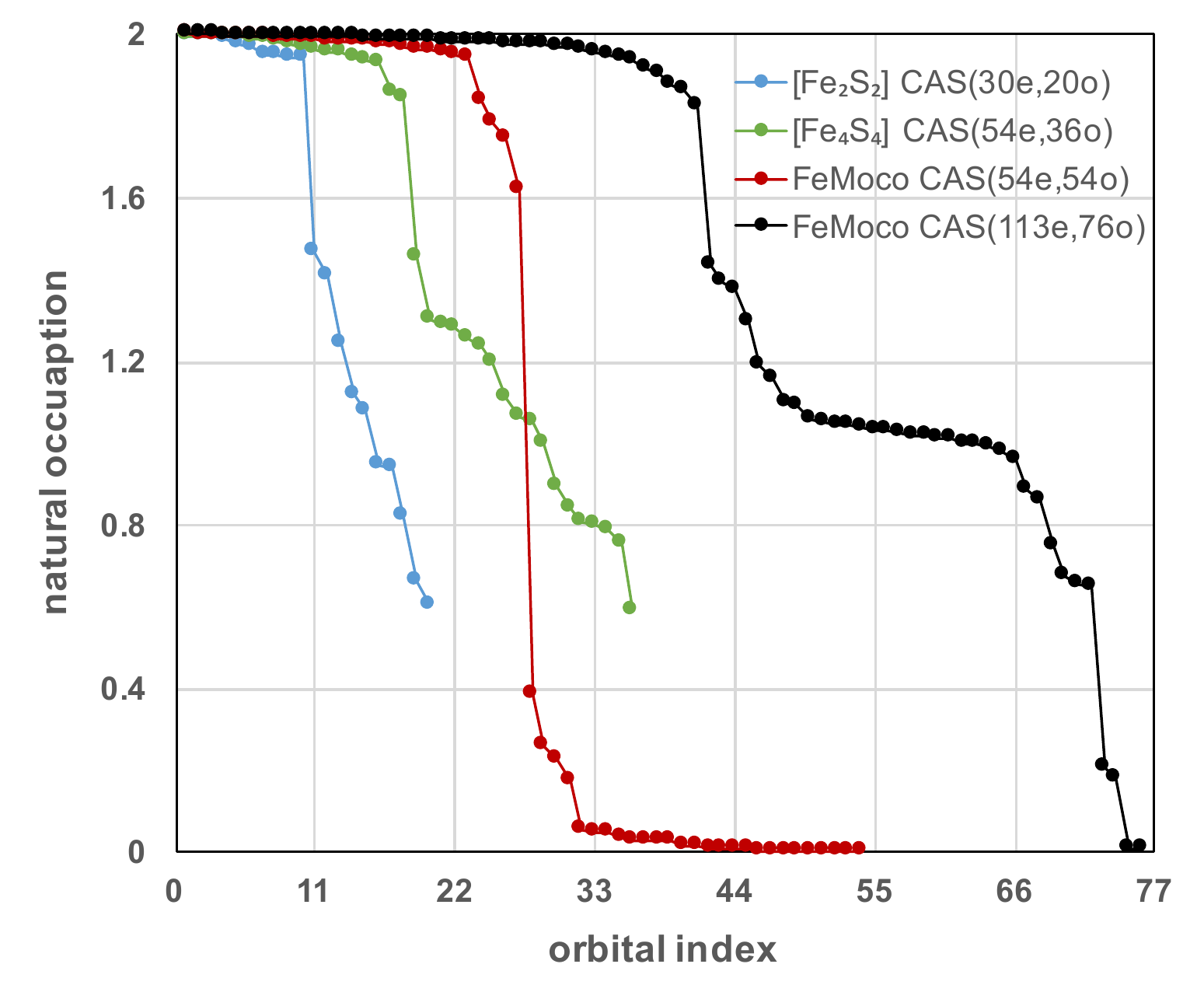}
\caption{Natural occupations obtained with DMRG for
$S=0$ state of a [\ce{Fe2S2}] complex with CAS(30e,20o) and $D=8000$,
$S=0$ state of a [\ce{Fe4S4}] complex with CAS(54e,36o) and $D=4000$,
$S=0$ state of FeMoco with CAS(54e,54o) reported in Ref. ~\onlinecite{ReiWieSvoWecTro-PNAS-17}
and $D=2000$, and the $S=3/2$ state of FeMoco with CAS(113e,76o) constructed in this work and $D=2000$.
In contrast to the other models, the CAS(54e,54o) ground-state has no open shells.}
\label{fig:occs}
\end{figure}

As we have mentioned, the reason for the simplicity of the classical simulations is not from the intrinsic electronic structure of the
FeMo cofactor but due to the active space in Ref.~\onlinecite{ReiWieSvoWecTro-PNAS-17}. In FeMoco, the Fe and Mo ions are expected to be in the (II), (III), or (IV) formal oxidation states\cite{munck1975nitrogenase,zimmermann1978nitrogenase,yoo2000mossbauer,bjornsson2014identification,spatzal2016nitrogenase,bjornsson2017revisiting},
which leads to approximately 35 open shells (singly filled orbitals) depending on the charge state of the cluster. The prevalence of Fe(II) and Fe(III) oxidation states
is supported experimentally by the Fe M\"ossbauer spectrum\cite{munck1975nitrogenase,zimmermann1978nitrogenase,yoo2000mossbauer},
and can be seen in direct theoretical calculations
of smaller pieces of the FeMo cofactor, such as the [\ce{Fe2S2}] or [\ce{Fe4S4}] clusters~\cite{sharma_low-energy_2014,spmps2017}.
However, the one-body density matrix in the FeMo cofactor model of Ref.~\onlinecite{ReiWieSvoWecTro-PNAS-17} has no open shells, as seen
from the eigenvalues of the one-body density matrix (Figure~\ref{fig:occs}).
A related point is that the coefficient of the dominant (natural orbital) determinant in SHCI is very large (0.67) 
indicating that the wavefunction has mainly single or few determinantal character, which
is not possible for a low-spin system with many open shells.
A large determinant weight has also been observed in Ref.~\onlinecite{TubMejEpsHaiLevHugJiaMcCBabHeaWha-ARX-18}
(in fact they observed an even
 larger determinant weight, probably because of using a smaller number of variational determinants).
As shown in Figure \ref{fig:energies}, the CCSD(T) energy is also within a few m$E_h$ of the variational DMRG
and extrapolated SHCI energies, confirming the single reference nature of this problem.

Although the electronic structure of the ground state within the active space of Ref.~\onlinecite{ReiWieSvoWecTro-PNAS-17} is qualitatively incorrect,
we nonetheless believe
that the quantum resource estimates in Ref.~\onlinecite{ReiWieSvoWecTro-PNAS-17}, e.g. for a Trotter step, that are the primary focus of the paper,
are probably reasonable and the main conclusions in that work are thus unaffected. This is because
%
the cost of the Trotter step relies primarily on the magnitude and number of the Hamiltonian matrix elements which does not vary much with different
choices of valence active space of similar size.
The character of the ground-state affects the efficiency of adiabatic state preparation in the quantum algorithm,
but this is left as an open problem in Ref.~\onlinecite{ReiWieSvoWecTro-PNAS-17}.

\begin{figure}[htb]
  \includegraphics[width=0.4\textwidth]{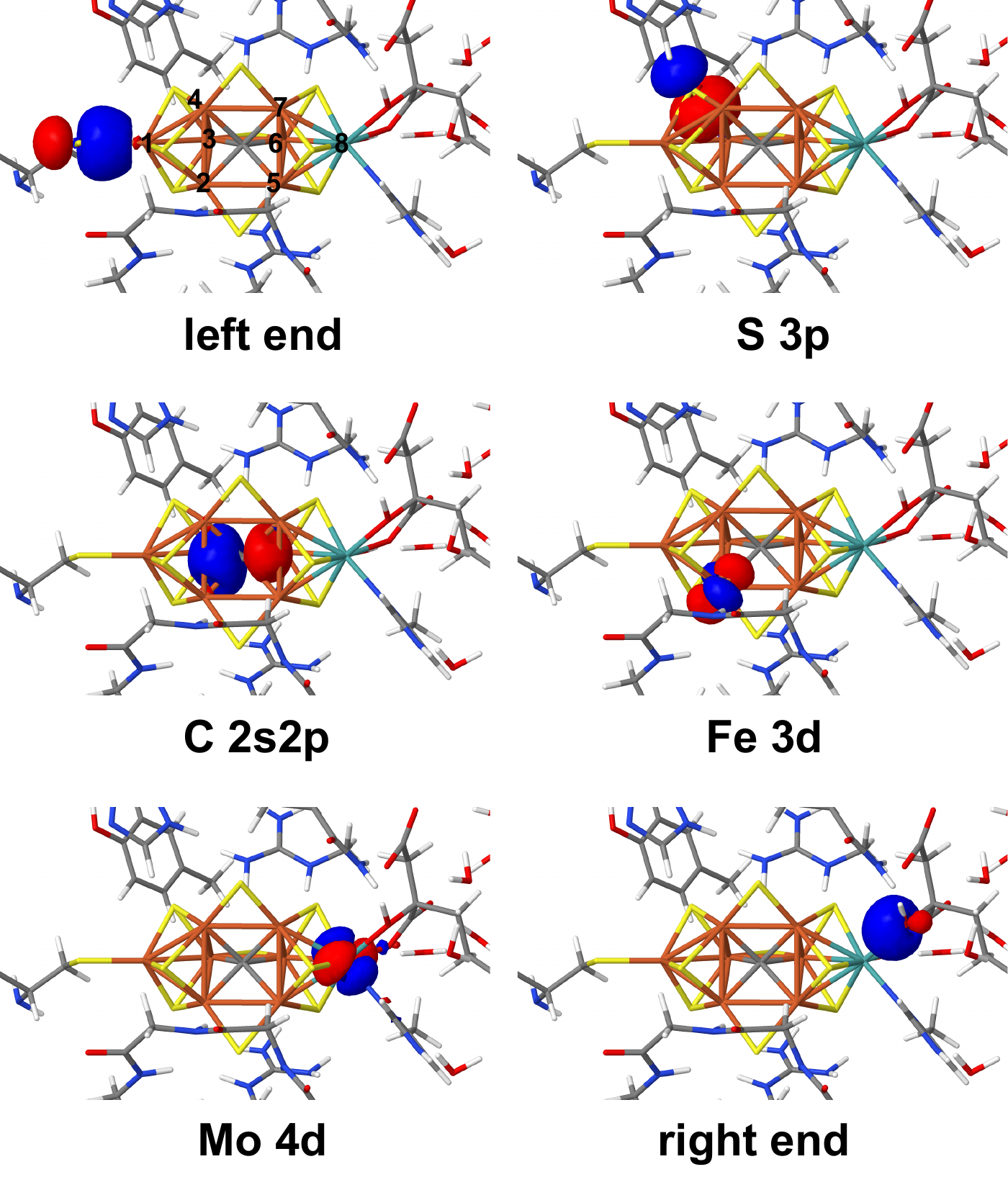}
  \caption{Illustration of some selected active orbitals for FeMoco
  in the active space model CAS(113e,76o) constructed in this work.}\label{fig:orbitals}
\end{figure}

Nonetheless, it seems desirable to have a more qualitatively reasonable active space for future studies.
For this purpose, we attach a valence active space Hamiltonian\cite{linkToFCIDUMP} of the FeMo cofactor constructed from all Fe 3d, S 3p, Mo 4d, and C 2s2p
orbitals in the [\ce{MoFe7S9C}] core, as well as some bonding ligand orbitals.
The active orbitals were obtained by first performing high-spin unrestricted Kohn-Sham calculations with the B3LYP functional\cite{becke1993density,lee1988development,stephens1994ab}
and the TZP-DKH\cite{jorge2009contracted} basis for Fe, S, and Mo,
and the def2-SVP basis\cite{weigend2005balanced}
for the other atoms (C, H, O, and N) using a structure in Ref. \onlinecite{bjornsson2017revisiting},
and then split-localizing the unrestricted natural orbitals.
This results in an active space model with 113 electrons in 76 orbitals.
The detailed composition is shown in Table \ref{tab:act}
and some selected localized orbitals are shown in Figure \ref{fig:orbitals}.
The dimension of the full configuration interaction (FCI) space is on the order of $O(10^{35})$
for the spin $S=3/2$ ground state\cite{munck1975nitrogenase,zimmermann1978nitrogenase} in this FeMoco active space.
We have performed preliminary DMRG calculations to check the qualitative features of the active space.
As shown in Figure \ref{fig:occs}, the natural occupation numbers
obtained with a DMRG solution ($D=2000$) for $S=3/2$ show a large number of
singly occupied orbitals, which demonstrates that
this active space captures the open-shell character of FeMoco
in sharp contrast with the previous model\cite{ReiWieSvoWecTro-PNAS-17}.
While we emphasize that a detailed and chemically meaningful study on FeMoco
should consider many other factors, such as the convergence of the environment representation, different protonations, etc.,
we conclude that the active space Hamiltonian we provide contains at least a qualitative model of the open-shell character
and low energy states of the cofactor.
We hope this will be useful in future quantum (or classical) estimates of the complexity of  FeMo cofactor electronic structure.

\textit{Acknowledgements}. ZL and GKC were supported by the US National Science Foundation via CHE-1665333.
JL and CJU were supported by AFOSR grant FA9550-18-1-0095.

\begin{table}[ht]
\caption{Composition of the active space with 76 orbitals for FeMoco.}\scriptsize
\begin{tabular}{ccl}
\hline\hline
group & orbital & orbital index \\
\hline
\multicolumn{3}{c}{left cubane} \\
1 & left end & 1, 2  \\
2 & Fe1 3d & 3, 4, 5, 6, 7 \\
3 & S 3p & 8, 9, 10, 11, 12, 13, 14, 15, 16 \\
4 & Fe2 3d & 17, 18, 19, 20, 21 \\
5 & Fe3 3d &  22, 23, 24, 25, 26 \\
6 & Fe4 3d & 27, 28, 29, 30, 31 \\
\multicolumn{3}{c}{central part} \\
7 & S 3p, C 2s2p  & 32, 33, 34, 35, 36, 37, 38, \\
  &                  & 39, 40, 41, 42, 43, 44 \\
\multicolumn{3}{c}{right cubane} \\
8 & Fe5 3d & 45, 46, 47, 48, 49 \\
9 & Fe6 3d & 50, 51, 52, 53, 54 \\
10 & Fe7 3d & 55, 56, 57, 58, 59 \\
11 & S 3p & 60, 61, 62, 63, 64, 65, 66, 67, 68 \\
12 & Mo8 4d & 69, 70, 71, 72, 73 \\
13 & right end & 74, 75, 76 \\
\hline\hline
\end{tabular}\label{tab:act}
\end{table}

\bibliographystyle{apsrev4-1}

%

\end{document}